\documentclass[preprint]{aastex631}
\usepackage{todonotes}

\shorttitle{LISA in a Galactic Background}
\shortauthors{Digman and Cornish}
\graphicspath{{./}{figures/}}

\begin{document}

\title{LISA Gravitational Wave Sources in A Time-Varying Galactic Stochastic Background}

\author[0000-0003-3815-7065]{Matthew C. Digman}
\affiliation{eXtreme Gravity Institute\\
Department of Physics\\
Montana State University\\
Bozeman, Montana 59717, USA}

\author[0000-0002-7435-0869]{Neil J. Cornish}
\affiliation{eXtreme Gravity Institute\\
Department of Physics\\
Montana State University\\
Bozeman, Montana 59717, USA}

\begin{abstract}

A unique challenge for data analysis with the Laser Interferometer Space Antenna (LISA) is that the noise backgrounds from instrumental noise and astrophysical sources will change significantly over both the year and the entire mission. Variations in the noise levels will be on time scales comparable to, or shorter than, the time most signals spend in the detector's sensitive band. The variation in the amplitude of the galactic stochastic GW background from galactic binaries as the antenna pattern rotates relative to the galactic center is a particularly significant component of the noise variation. LISA's sensitivity to different source classes will therefore vary as a function of sky location and time. The variation will impact both overall signal-to-noise and the efficiency of alerts to EM observers to search for multi-messenger counterparts. 

\end{abstract}



\section{Introduction} \label{sec:intro}
The Laser Interferometer Space Antenna (LISA) is a space-based mHz gravitational wave (GW) observatory set to launch in the mid-2030s \cite{2017arXiv170200786A}. Perhaps hundreds of millions of galactic ultra-compact binaries (UCBs), primarily white dwarf binaries, will continuously contribute to the gravitational-wave signal seen by LISA, with the brightest sources likely being first detected within days or even hours of LISA entering observing mode \citep{Cornish:2017vip,Timpano:2005gm}. Over the mission's lifetime, LISA will likely detect tens of thousands of these sources \citep{Cornish:2017vip}. Dozens of LISA-detectable white dwarf binaries have already been detected electromagnetically \citep{Stroeer:2006rx,2018MNRAS.480..302K}, and are currently the only guaranteed-in-advance sources for multi-messenger gravitational-wave astronomy. Such sources, which are bright in both the electromagnetic (EM) spectrum and gravitational waves, present excellent opportunities for white dwarf science \citep{2011CQGra..28i4019M}, tests of general relativity \citep{Littenberg:2018xxx}, and checking the instrumental calibration of LISA \citep{2018PhRvD..98d3008L}. 

While the enormous quantity of galactic GW sources is a scientific opportunity for LISA, it also presents a challenge for LISA's other observations. The galaxy contains far more GW sources than LISA will be able to resolve. These unresolved sources will combine to form a stochastic gravitational-wave background which will compete with other expected LISA science sources, such as Supermassive Black Hole Binaries (SMBHBs), Stellar Origin Black Hole Binaries (SOBHBs), and Extreme Mass Ratio Inspirals (EMRIS). The stochastic background from galactic sources could also mask extragalactic and cosmological stochastic backgrounds \citep{Bonetti:2020jku,Boileau:2021sni,Adams:2013qma}, which could be a powerful test of models of early universe cosmology \citep{Auclair:2019wcv,Ricciardone:2016ddg}. 

To maximize the impact of LISA's transient source alerts to EM observers, the stochastic background must be characterized and updated continuously to give the best possible inputs to parameter constraint pipelines and mitigate the impacts on other source classes \citep{NASALISAStudyTeam:2020lee}. As new galactic sources are detected and characterized, they can be subtracted from the LISA data stream, reducing the amplitude of the stochastic background. Because the signals from galactic sources are continuous, subtracting them from the data will retrospectively improve constraints on past transient sources throughout the mission's lifetime and enhance the sensitivity of future alerts. 

Existing analyses have generally modeled the galactic stochastic background as constant throughout the year \citep{Timpano:2005gm,Karnesis:2021tsh}. In reality, LISA's quadrupolar antenna pattern continuously rotates across the sky, causing the sensitivity as a function of sky location to vary over time. The rotating sensitivity would still produce an approximately constant noise level for an isotropic stochastic signal (as expected from cosmology). However, like most mass in the galaxy, GW emission from galactic binaries is expected to be highly concentrated in the direction of the galactic center \citep{Cornish:2002bh}. As a result, the galactic stochastic gravitational-wave background amplitude will vary significantly throughout of the year based on the position of LISA's antenna pattern relative to the galactic center. In this analysis, we adopt an improved technique to better model the effect of this variation on LISA. 

This paper describes a time-frequency decomposition of the stochastic background with LISA. This decomposition allows us to effectively model the galactic stochastic background as a source of cyclostationary noise, consisting of a time-independent spectral contribution to the noise signal modulated by a frequency-independent amplitude. 

The paper is organized as follows. In Section.~\ref{sec:methods}, we adapt existing iterative fitting procedures to a time-varying galactic spectrum, and show that the decomposition results in a much-improved noise model compared to the previously assumed constant-noise model. In Section~\ref{ssec:fitting}, we provide coefficients for an empirical fitting formula to the cyclostationary spectrum. In Section~\ref{sec:results}, we explore the results of using the improved cyclostationary model on the detectability of galactic binaries and SMBHBs. In Section~\ref{sec:conclusion}, we draw conclusions and look toward our result's impact on future analysis. 

\section{Methods} \label{sec:methods}

To develop the time-varying galactic stochastic background, we use the catalog of galactic binary sources from the ``Sangria'' dataset produced from a population synthesis model for the LISA data challenge \citep{Baghi:2022ucj}. Other galactic populations could produce different results \citep{Nissanke:2012eh,Lamberts:2019nyk,Breivik:2019lmt}. To generate a model for the noise background as a function of time, we use a version of the iterative fitting procedure described in \citep{Timpano:2005gm} and \citep{Karnesis:2021tsh} adapted to a time-varying spectrum. See \citep{Adams:2013qma} for a different approach to handling the temporal variation in the galactic spectrum. 

Calculating the signals in the time-frequency domain allows the methodology to be efficiently adapted to handle a time-varying spectrum. We use the Wilson-Daubechies-
Meyer (WDM) wavelet basis \citep{Necula:2012zz} with a modified normalization chosen such that a unit power spectrum in the wavelet domain $S_{nm}=1$ generates unit variance white noise in the time domain. The basis has a uniform tiling in frequency and time, which permits generating a uniform grid of frequency spectra as a function of time. 

We generate LISA waveforms as described in \cite{2020PhRvD.102l4038C} using the Rigid Adiabatic approximation \citep{Cornish:2020vtw,Rubbo:2003ap}

We first generate a realization of the signal seen by LISA as

\begin{equation}\label{eq:lisa_wavelet_signal0}
w_{nm}=w_{nm,inst}+\sum_k{w_{nm,k}},
\end{equation}

where $w_{nm,inst}$ is a random realization of the instrumental noise and $w_{nm,k}$ are the wavelet decompositions of each of the 29,000,036 binaries in the Sangria dataset. The vast majority of these binaries produce too faint a signal to be individually detectable. To mitigate unnecessary evaluations, we do an initial pass where we calculate the instrumental-noise-only SNR and add every source with $SNR_{inst}<SNR_{thresh}$ into an 'irreducible' background $w_{nm,irr}$, so that we can rewrite:

\begin{equation}\label{eq:lisa_wavelet_signal1}
w^{AET}_{nm}=w^{AET}_{nm,inst}+w^{AET}_{nm,irr}+\sum_{k'}{w^{AET}_{nm,k'}},
\end{equation}

where now the sum over $k'$ runs only over the $\mathcal{O}(100,000)$ binaries bright enough to pass the initial minimum SNR cut. This component of the procedure will not be possible for the real LISA data because we will be starting from the observed $w^{AET}_{nm}$ and not an artificially simulated catalog. To further accelerate convergence, we also remove the bright binaries with $SNR_{inst}>30$ at this step. At a later iteration, we verify all removed bright binaries retain $SNR_{calc}>SNR_{thresh}$ and replace them if necessary.

\subsection{The Cyclostationary Model}\label{ssec:cyclostationary}

From Eq.~\ref{eq:lisa_wavelet_signal1}, we must estimate the noise level $S^{AE}_{nm}$ for the two channels which have an appreciable component from the galactic stochastic background. As a first approximation, we can approximate the mean noise spectrum as: 
\begin{equation}\label{eq:lisa_wavelet_SAE_mean}
S^{AE}_{m} = \frac{1}{N_t}\sum_{n=0}^{N_t}\left(w^{AE}_{nm}\right)^2.
\end{equation}

We assume the instrumental component is constant in time and the same for both channels, although relaxing this assumption would be straightforward in the time-frequency domain. \\
To write the signal as cyclostationary, we decompose the signal into two parts, one depending only on time and one only on the frequency:

\begin{equation}\label{eq:lisa_cyclostationary_base}
S^{AE}_{nm,cyclo}=r^{AE}_n\langle S_{m,gal}\rangle+S_{m,inst},
\end{equation}
where $\langle S_{m,gal}\rangle$ is the mean galactic spectrum averaged over time and the A and E channels, and $r^{AE}_n$ is a multiplier which varies in time and will be different for the A and E channels. To determine $r^{AE}_n$, we can average over frequency

\begin{equation}\label{eq:r_cyclostationary}
r^{AE}_n = \frac{1}{m_2-m_1}\sum_{m'=m_1}^{m'=m_2} \frac{S^{AE}_{nm'}-S_{m',inst}}{\langle S_{m',gal}\rangle},
\end{equation}

where the sum from $m_1$ to $m_2$ runs over only pixels where $\langle S_{m,gal}\rangle/\langle S_{m,instr}\rangle >5$ to reduce the influence of numerical noise on $r^{AE}_n$. 

For the particular test case presented in this paper, we can separate the instrumental noise from the galactic signal exactly, so it is possible to instead obtain a more precise fit to $r^{AE}_n$ by subtracting the instrumental signal exactly from $w^{AE}_{nm}$ to obtain instead

\begin{equation}\label{eq:r_cyclostationary_alt}
r^{AE}_n = \frac{1}{m_2-m_1}\sum_{m'=m_1}^{m'=m_2} \frac{S^{AE}_{nm',gal}}{\langle S_{m',gal}\rangle},
\end{equation}

where $m_1$ and $m_2$ can now obtain a numerically stable average over a larger range in frequency, for example $\langle S_{m,gal}\rangle/\langle S_{m,instr}\rangle > 0.1$. Although this version of the procedure would not be possible for real LISA data, this more precise estimator of $r^{AE}_n$ allows us to more thoroughly investigate the limitations of the cyclostationary model compared to a hypothetical perfectly optimal model of the galactic stochastic background. 

\begin{figure}
\includegraphics[width=0.8\columnwidth]{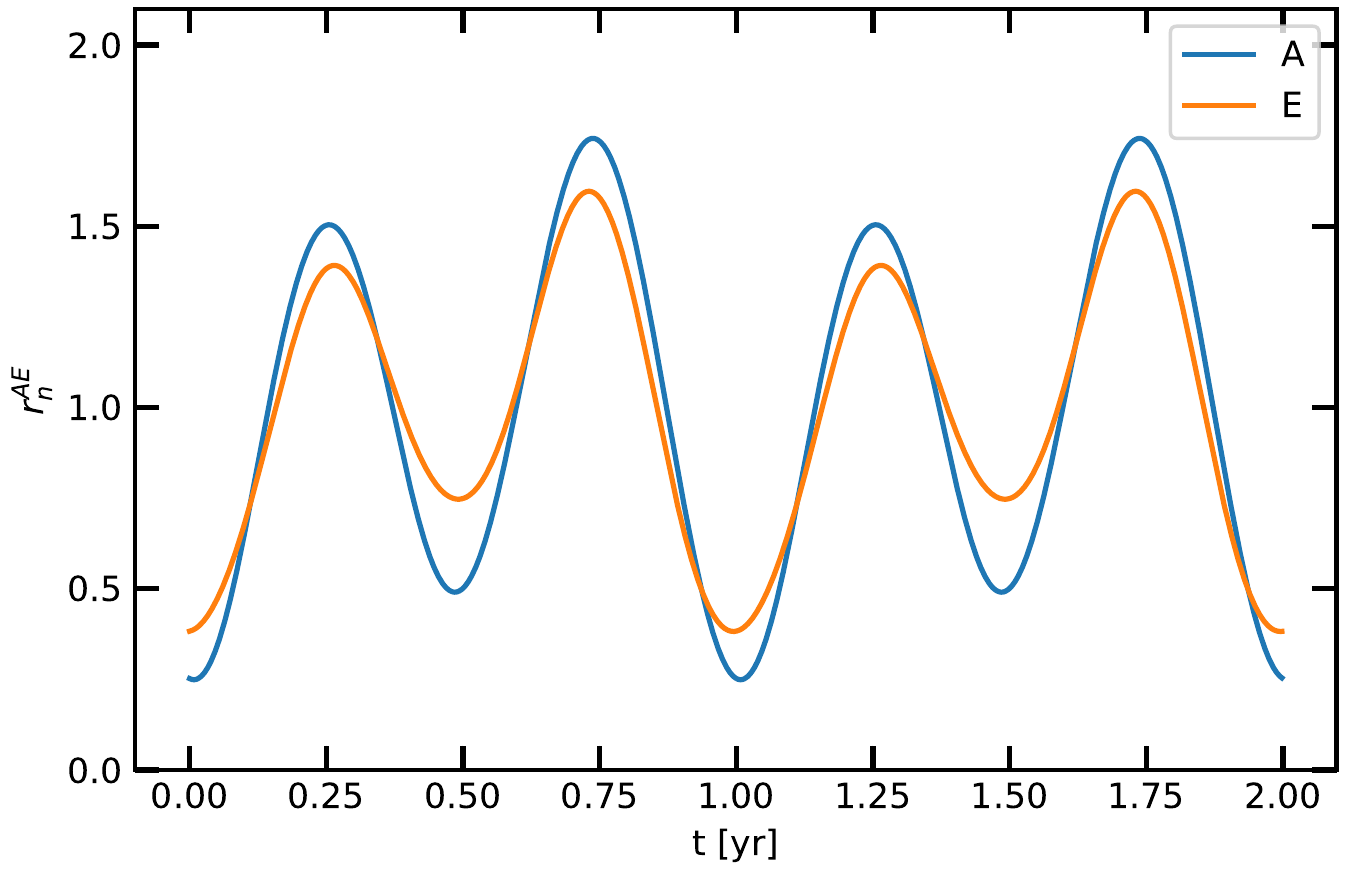}
\caption{Modulation in the amplitude of the galactic background $r^{AE}_n$ from Eq.~\ref{eq:r_cyclostationary_alt} as a function of time for two years of simulated LISA data. The amplitude in the A channel varies by a factor of $\simeq7$, while the amplitude in the E channel varies by a factor of $\simeq4$, and the mean of the envelopes varies by a factor of $\simeq5$, from $\simeq0.32$ to $\simeq1.67$. The model includes the first 5 harmonics of $1.\;\text{yr}$ as described in Eq.~\ref{eq:r_periods}, although in this case the amplitudes of the first two harmonics dominate the other three by more than an order of magnitude.}\label{fig:gb_amp_envelope}
\end{figure}

Eq.~\ref{eq:r_cyclostationary} allows for an arbitrary cyclostationary noise model. Physically, the time variation of the signal is generated entirely by the shifting antenna pattern due to the rotation of the constellation relative to the galaxy throughout the year \citep{Cornish:2001hg}. Consequently, we can generate a smoother noise model by extracting only the frequency components in harmonics of 1 year:
\begin{equation}\label{eq:r_periods}
r^{AE}(t) \simeq 1+\sum_{k}{A_{AE,k}\cos\left(2\pi k t/T_{year}-\phi_{AE,k}\right)},
\end{equation}

where $r^{AE}_n=r^{AE}(t_n)$. The detailed distribution of the population of galactic binaries will ultimately determine the number of harmonics necessary to fit the time evolution of the amplitude, but there are practical limits to LISA's ability to place constraints on higher harmonics \citep{Edlund:2005ye,Renzini:2021iim,Nissanke:2012eh,Cornish:2002bh}. We find that the first five harmonics of 1 year, $k=1...5$, are sufficient to capture essentially all of the physical time variation of the signal while keeping the spurious variation in $r^{AE}_{n}$ due to noise artifacts minimal. The coefficients $A_{AE,k}$ and $\phi_{AE,k}$ can be extracted from either the Fast Fourier Transform of $r^{AE}_n$ or a least-squares fit. A fit to $r^{AE}_n$ for our two year simulated dataset is shown in Fig.~\ref{fig:gb_amp_envelope}, and parameters for several different observing time intervals are shown in Table.~\ref{tab:evolve_params}.

\begin{center}
\begin{table}
\begin{tabular}{||c c c c c c c c c c c c||} 
 \hline
$T_{obs}$ [yr] & Channel & $A_1$ & $\phi_1$ & $A_2$ & $\phi_2$ & $A_3$ & $\phi_3$ & $A_4$ & $\phi_4$ & $A_5$ & $\phi_5$\\ [0.5ex]
 \hline\hline
1 & A & 0.183 &  3.92 & 0.616 &  3.09 & 0.012 &  4.92 & 0.004 &  3.33 & 0.005 &  4.72\\
\hline
1 & E & 0.212 &  3.56 & 0.462 &  3.08 & 0.022 &  0.94 & 0.027 &  0.08 & 0.006 &  1.84 \\
\hline
2 & A & 0.177 &  3.92 & 0.622 &  3.10 & 0.012 &  4.93 & 0.003 &  3.83 & 0.004 &  4.49 \\
\hline
2 & E & 0.211 &  3.54 & 0.458 &  3.08 & 0.023 &  0.96 & 0.023 &  0.05 & 0.004 &  2.01 \\
\hline
4 & A & 0.181 &  3.91 & 0.625 &  3.09 & 0.016 &  5.38 & 0.006 &  3.98 & 0.004 &  4.20 \\
\hline
4 & E & 0.209 &  3.58 & 0.462 &  3.08 & 0.022 &  1.23 & 0.023 &  0.03 & 0.002 &  1.11 \\
\hline
8 & A & 0.183 &  3.95 & 0.630 &  3.09 & 0.016 &  5.47 & 0.008 &  3.85 & 0.005 &  4.38 \\
\hline
8 & E & 0.207 &  3.58 & 0.467 &  3.08 & 0.023 &  1.12 & 0.024 &  0.05 & 0.003 &  1.74 \\
 \hline
\end{tabular}
\caption{\label{tab:evolve_params}Amplitudes and phases  for use in Eq.~\ref{eq:r_periods}, computed from the Fast Fourier Transform of $r^{AE}_n$ computed using the spectra shown in Fig.~\ref{fig:gb_evolve} at several sample observation times. The first two harmonics dominate in all cases.}
\end{table}
\end{center}

\subsection{The Iterative Fit}\label{ssec:iterative}
The objective of the iterative fitting procedure is to remove all binaries bright enough that their parameters would be sufficiently well characterized by a full MCMC pipeline that their signal could be subtracted from the LISA data stream. Because the binaries interfere with each other's signals, subtracting a bright binary will often cause further fainter binaries to become detectable in the data stream. We use the well-established iterative procedure described in, e.g., \cite{Karnesis:2021tsh}. 

By iterating this procedure until there are no bright binaries left, we arrive at an estimate of the noise level as a function of time $S^{AE}_{nm}$. For this analysis, we choose a cutoff of $\text{SNR}_{thresh}=7$ as the threshold for a galactic binary to be considered detectable. Our method allows directly comparing results between a more realistic cyclostationary noise model, described in Sec.~\ref{ssec:cyclostationary}, to the constant noise model commonly used in previous analyses, equivalent to setting $r^{AE}_n=1$ in Eq.~\ref{eq:r_cyclostationary}. Because the incoherent superposition of signals becomes intrinsically reasonably smooth after a few iterations, we adopt a Gaussian smoothing with a smoothing length that decreases significantly over the first few iterations, rather than the simple moving average or windowed median smoothing used in \cite{Karnesis:2021tsh}. The smoothing procedure is described in more detail in Sec.~\ref{ssec:smoothing}. 

For approximately the first two iterations of the iterative subtraction procedure, the signal is not well approximated as cyclostationary due to distortion from the individual influence of bright, well-isolated galactic binaries. For those iterations, we set $r^{AE}_{n}=1$. To prevent the initial iterations with a different noise model and smoothing length from causing sources to be incorrectly subtracted, we use a more conservative $\text{SNR}_{thresh}$ for those first few iterations. In testing, an exponential phase-in of the SNR cutoff from the initial $\text{SNR}_{thresh}=7(1+e^{-2k+4})$  for iterations $k=1..3$ and $\text{SNR}_{thresh}=7$ thereafter works well to prevent any sources from ever being incorrectly subtracted due to the changing smoothing length or time dependence, while ensuring each iteration subtracts almost as many sources as possible. As long as the cutoff is sufficiently conservative to prevent any erroneous subtractions of sub-threshold sources, the exact details of the phase-in of the cutoff should not affect the final result at all. However, the rate of phase-in does affect the number of iterations and total compute time it takes to achieve the final converged result.

\begin{figure}
\includegraphics[width=0.8\columnwidth]{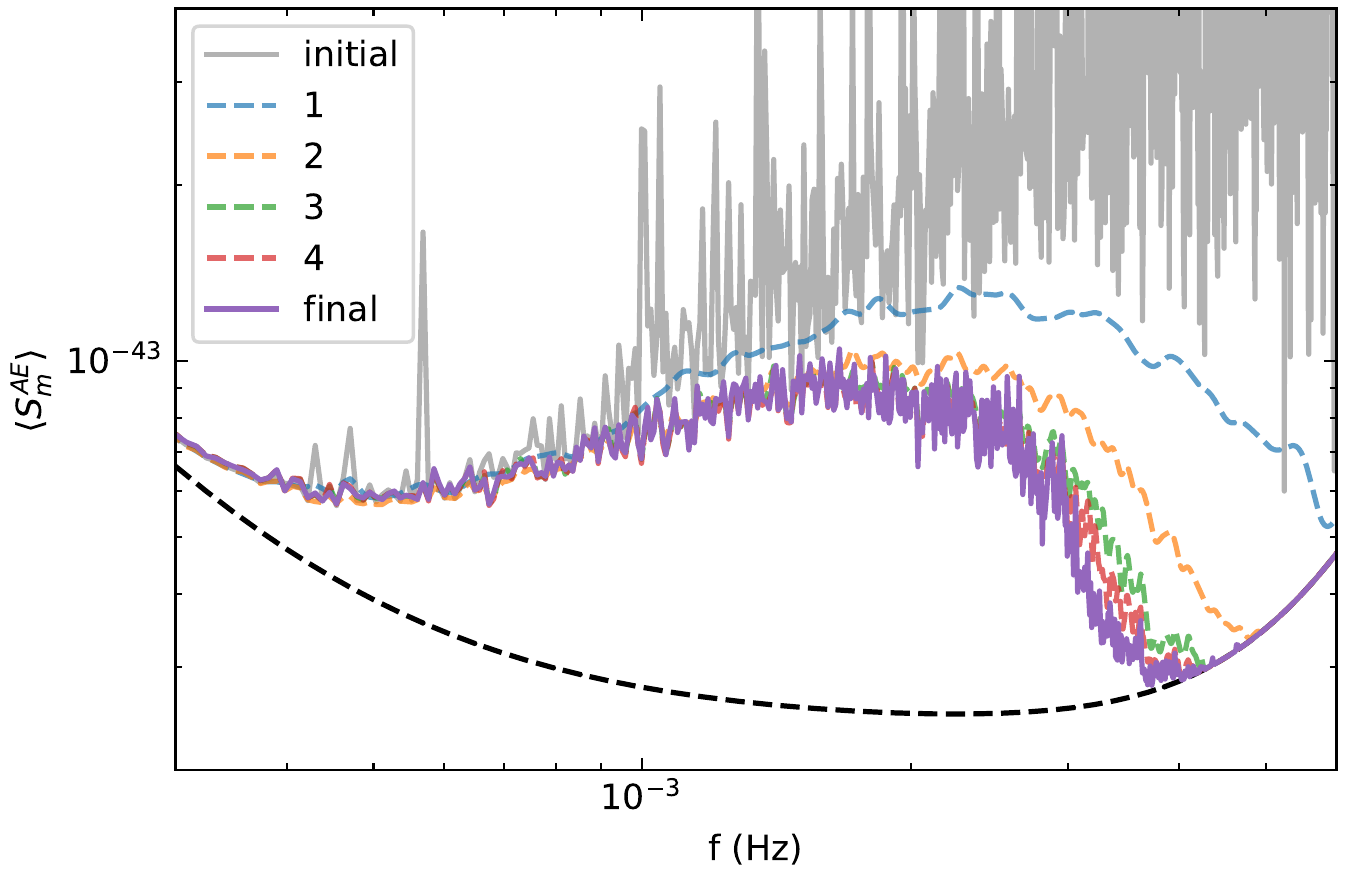}
\caption{Convergence of the iterative procedure described in Sec.~\ref{ssec:iterative} for a two year galactic stochastic gravitational-wave spectrum. The the final converged spectrum was first reached on iteration 17; iterations 5-16 are omitted to improve the clarity of the plot. The procedure has already suppressed 90\% of $\text{SNR}>7$ binaries by iteration 4, and the final 7 iterations are spent deciding on the inclusion of only $\sim10$ binaries with $\text{SNR}\simeq7$. \label{fig:gb_iter_converge}}
\end{figure}

An example illustrating the convergence to a galactic spectrum for two years of data using our procedure is shown in Fig.~\ref{fig:gb_iter_converge}. In this example, the procedure has essentially converged after $<10$ iterations, and reached full convergence on iteration 17. 

\subsection{Smoothing}\label{ssec:smoothing}
During the iterative fitting procedure, it is necessary to smooth the modeled galactic background in frequency so that individual bright binaries do not artificially compete with their own power. The final output power spectrum does not include any smoothing. To accomplish the smoothing, we apply a small Gaussian smoothing in log-frequency to the whitened log-power spectrum,    

\begin{equation}\label{eq:lisa_wavelet_SAE_smooth0}
\langle S^{AE}_{m,gal,smooth}\rangle =  \exp\left(G_{\sigma(\log(f))}*\log\left(\langle S_{m,gal}\rangle+\epsilon\right)\right)-\epsilon,
\end{equation}

where the width of the Gaussian smoothing in log-frequency is chosen such that, for the first iteration, the width around the peak at 2 mHz is  $\sigma(\log(2\;\text{mHz}))=6$ frequency pixels. Because the smoothing is only really necessary for the first few iterations, we reduce the smoothing length by one e-folding per iteration to a floor of $\sigma(\log(2\;\text{mHz}))=0.25$ pixels starting in the fifth iteration.

The addition and subsequent subtraction of the factor of $\epsilon=10^{-50}$ is a small numerical stabilizer to prevent the argument of the logarithm from becoming zero or negative, and is chosen to be much smaller than the lowest value of $\langle S_{m,gal}\rangle$ of interest. Using $\langle S_{m,instr}\rangle$ as the numerical stabilizer would obtain a more realistic cutoff. However, for the purposes of this paper, the smaller numerical stabilizer complements the more precise estimator of $r^{AE}_{n}$ in Eq.~\ref{eq:r_cyclostationary_alt} to allow us to investigate the ultimate limitations of the cyclostationary model to the best possible precision. 

We can then replace  $\langle S_{m,gal}\rangle$ with $\langle S^{AE}_{m,gal,smooth}\rangle$ in Eq.~\ref{eq:lisa_cyclostationary_base} throughout the iterative fitting procedure.

\section{Empirical Fitting Formula}\label{ssec:fitting}
The smoothed average frequency spectrum in Eq.~\ref{eq:lisa_wavelet_SAE_smooth0} allows the spectrum fitting to have essentially monotonic convergence to a final self-consistent estimate of the spectrum of the galactic background and is agnostic regarding the predicted frequency distribution of galactic sources. However, for some applications, it is desirable to reduce the number of parameters in the fit and obtain a smoother spectrum by fitting to a spectrum of known shape. A variety of ways of parameterizing the shape of the galactic background exist \citep{Flauger:2020qyi,Caprini:2019pxz,Karnesis:2021tsh}. Similar to \cite{Karnesis:2021tsh}, we use a 5-parameter model of the galactic background: 
\begin{equation}\label{eq:gal_evolve}
S_{gal}(f)=\frac{Af^{5/3}}{2}e^{-(f/f1)^\alpha}(1+\tanh((f_{knee}-f)/f_2)),
\end{equation}

where $\log_{10}(f_{knee})=a_k\log_{10}(T_{obs})+b_k$ and $\log_{10}(f_1)=a_1\log_{10}(T_{obs})+b_1$ can be used to predict the coefficients as a function of observation duration, where $T_{obs}$ is in years. The improvement in the overall amplitude of the galactic stochastic background as a function of $T_{obs}$ is absorbed by the variation in $f_1$, so a shift in $A$ as a function of time is not necessary. For fitting purposes $A$ and $f_2$ have a large dynamic range, so we search over the logarithms $\log_{10}A$ and $\log_{10} f_2$ instead, and fit over the spectrum in the wavelet domain, $\langle S_{m,gal}\rangle\simeq S_{gal} (f)/dt$.

To fit the parameters of Eq.~\ref{eq:gal_evolve}, we calculate the galactic spectra $\langle S^{AE}_{m,gal}(T_{\text{obs}})\rangle$ in 1 year increments $T_{\text{obs}}=1..8\;\text{yr}$ using the iterative fitting procedure described in Sec.~\ref{ssec:iterative}. We then use the dual annealing algorithm described in \cite{dual_annealing} as implemented in SciPy \citep{2020SciPy-NMeth} to obtain a least-squares fit of $\log_{10}\left(\langle S^{AE}_{m,gal}(T_{\text{obs}})\rangle+S_{m,inst}\right)$ to the 7-parameter model in Eq.~\ref{eq:gal_evolve}. We fit both channels and all eight years of data simultaneously. 

\begin{figure}
\includegraphics[width=0.8\columnwidth]{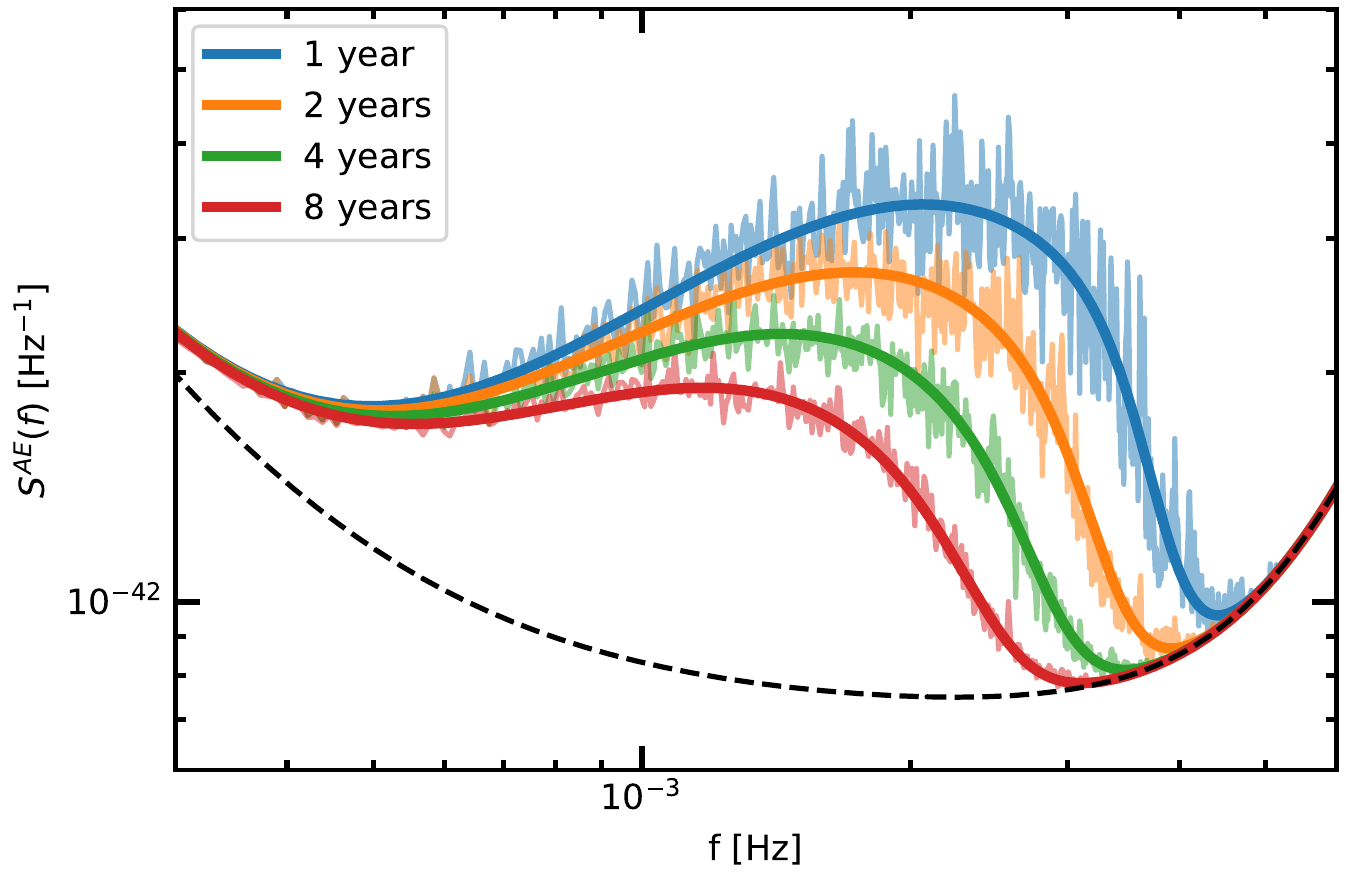}
\caption{Improvement of galactic confusion noise spectrum with observation time in the $\text{SNR}_{\text{thresh}}=7$ cyclostationary model, with the best fit spectrum from Eq.~\ref{eq:gal_evolve}. Our fits are all performed intrinsically in the wavelet domain, but over this frequency range we approximate $S^{AE}(f)\simeq\langle S^{AE}_m\rangle dt$, $dt\simeq30.075\;s$, to enable easier comparison to frequency domain results. Overall the model produces an excellent fit to the overall shape of the spectrum.}\label{fig:gb_evolve}
\end{figure}

The empirical best fit parameters are given in Table~\ref{tab:fit_params}, with spectra plotted for several selected observing times in Fig.~\ref{fig:gb_evolve}. The best fit spectral parameters shown in Table~\ref{tab:fit_params} are similar whether or not we use the cyclostationary model, and are in good general agreement with the results in \cite{Karnesis:2021tsh} for all the shape parameters, given expected differences due to our different smoothing and fitting procedures. Our amplitude normalization in the wavelet domain is chosen such that $S_{nm}=1$  corresponds to unit variance white noise in the time domain for a grid size $N_f=2048$, $N_t=512\;\text{pixels}/\text{year}$. To enable easier comparison to other frequency domain results, we report instead $A=dt A_{\text{wavelet}}$, where $dt\simeq30.075\;s$. 
In combination with the parameters in Table~\ref{tab:evolve_params} used in the formula for $r^{AE}_n$ in Eq.~\ref{eq:r_periods}, the parameters in Table~\ref{tab:fit_params} can be used to obtain a reasonable 25 parameter fit to our full cyclostationary model of the galactic spectrum fit to Sangria data for any given $T_{obs}$,
\begin{equation}\label{eq:r_ft}
S_{gal}(t,f)\simeq r^{AE}(t)S_{gal}(f).
\end{equation}
A full joint least-squares fit of the time and frequency parameters would likely find a slightly better best-fit model, but would be significantly more computationally expensive.

\begin{center}
\begin{table}
\begin{tabular}{||c c c c c c c c||} 
 \hline
Model & $a_1$&$a_k$&$b_1$&$b_k$&$A\times 10^{37}$&$ f_2$ [mHz]&$\alpha$\\ [0.5ex] 
 \hline\hline
Cyclostationary & -0.24 & -0.21 & -2.71 & -2.44 &2.27& 0.51 & 1.52\\
\hline
Constant &  -0.24&  -0.21&-2.70&-2.44&2.13& 0.53&  1.58\\
 \hline
\end{tabular}
\caption{\label{tab:fit_params}The best fit parameters to the fitting formula in Eq.~\ref{eq:gal_evolve} in both our cyclostationary model and a constant noise model. The shape parameters change only slightly between the models, while there is a small shift in the overall amplitude normalization. Note that the normalization of our amplitudes $A$ is different from \cite{Karnesis:2021tsh}. Although we fit the amplitudes to the mean spectrum in the wavelet domain, to enable easier comparison to other frequency domain results, we report instead $A=dt A_{\text{wavelet}}$, where $dt\simeq30.075\;s$. Our normalization in the wavelet domain is chosen such that for $N_f=2048$, $N_t=512\;\text{pixels}/\text{year}$, a unit power spectrum in the wavelet domain corresponds to unit variance white noise in the time domain. }
\end{table}
\end{center}

\section{Results}\label{sec:results}

\subsection{Improving The Fit}\label{ssec:white_noise}
\begin{figure}
\includegraphics[width=\columnwidth]{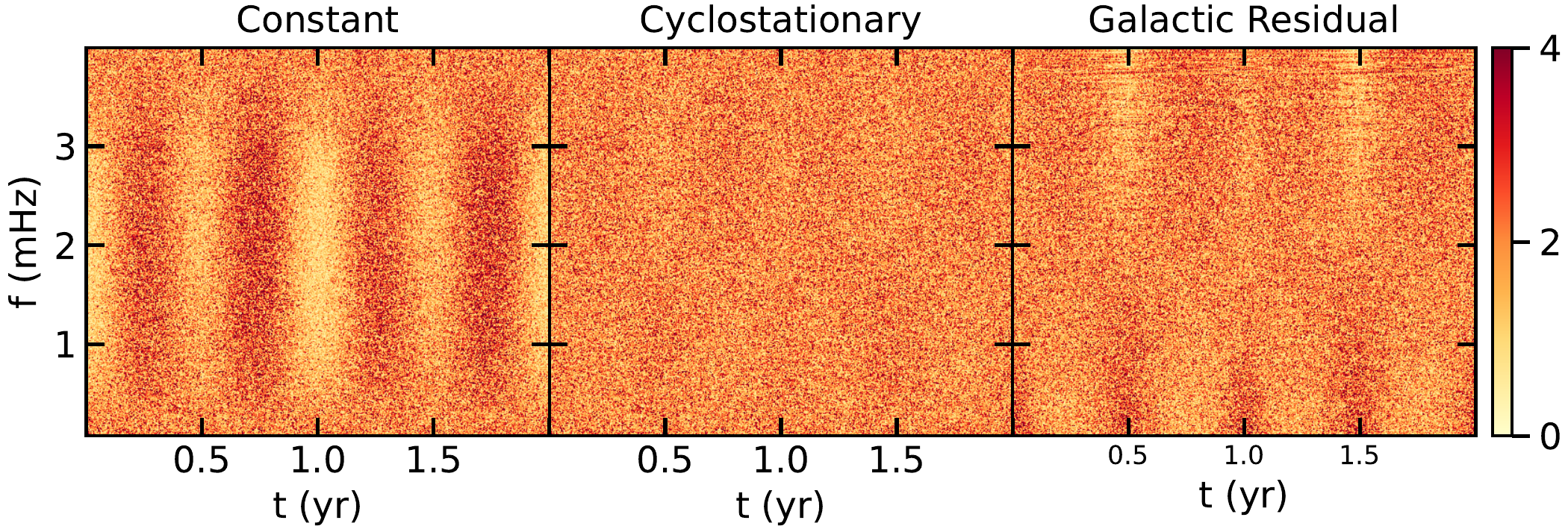}
\caption{\label{fig:whitened_residual}Comparison of whitened residual power $(w^{AE}_{nm})^2/S^{AE}_{nm,model}$ between \textbf{Left: }a constant model of the galactic stochastic background and \textbf{Center:} our cyclostationary model for a two year simulated dataset. The cyclostationary model is a significant improvement, with residuals statistically well approximated as Gaussian. In \textbf{Right:} we show the whitened residuals of the cyclostationary component after turning off instrumental noise completely, i.e. $(w^{AE}_{nm,gal})^2/S^{AE}_{nm,gal}$. At frequencies around $2\;\text{mHz}$, the signal is still very well approximated as cyclostationary, while at the tails of the spectrum, a slight deviation from perfect cyclostationarity is apparent. This deviation is of limited practical significance because the spectrum falls off rapidly compared to the instrumental noise at those frequencies. Additionally, it will not be possible to separate the instrumental and galactic contributions to the noise spectrum in real data}
\end{figure}

In this section, we evaluate quantitatively how the cyclostationary model improves the fit of the galactic stochastic background compared to a constant background model. To assess the improvement, we compare the whitened residuals from the simulated dataset; when the model is a good fit, the residuals will approach unit variance Gaussian white noise. This comparison is shown in Fig.~\ref{fig:whitened_residual}. 

We can then use an Anderson-Darling test \citep{doi:10.1080/01621459.1954.10501232,Jones1987EmpiricalPW} to detect deviations from normality. The constant model residuals give a test statistic $A^{*2}_{\text{const}}\approx37.24$, an extremely significant detection of deviation from normality ($p<10^{-5}$ corresponds to a test statistic $A^{*2}_{\text{const}}\simeq2.28$). The cyclostationary model residuals give $A^{*2}_{\text{const}}\approx0.30$, which is unable to reject the null hypothesis of normality at $p>0.1$. Both sets of residuals are consistent with zero mean and unit variance. 

Because we can separate the instrumental and galactic noise perfectly with our simulated ``Sangria'' dataset, we can investigate the true level of deviation from Gaussian white noise in the cyclostationary model, shown in the right panel of Fig.~\ref{fig:whitened_residual}. Although this plot makes a slight deviation from cyclostationarity apparent by eye, note that, even in the complete absence of instrument noise, the overall distribution of the residuals still gives $p>0.1$ consistency with normality by the Anderson-Darling test for $0.1\;\text{mHz}<f<4\;\text{mHz}$. 

The deviation is likely due to a combination of the slight frequency dependence of the LISA antenna pattern in this frequency range (see, e.g., \cite{Cornish:2002bh}) and the fact that there are fewer nearly-detectable binaries to average over at frequencies near the tails of the galactic spectrum. The deviation from the cyclostationary model is of little practical significance where the instrumental and galactic contributions to the noise cannot be perfectly separated, as in real data.

\subsection{Sensitivity to Galactic Binaries}\label{ssec:gal_sens}

\begin{figure}
\includegraphics[width=0.8\columnwidth]{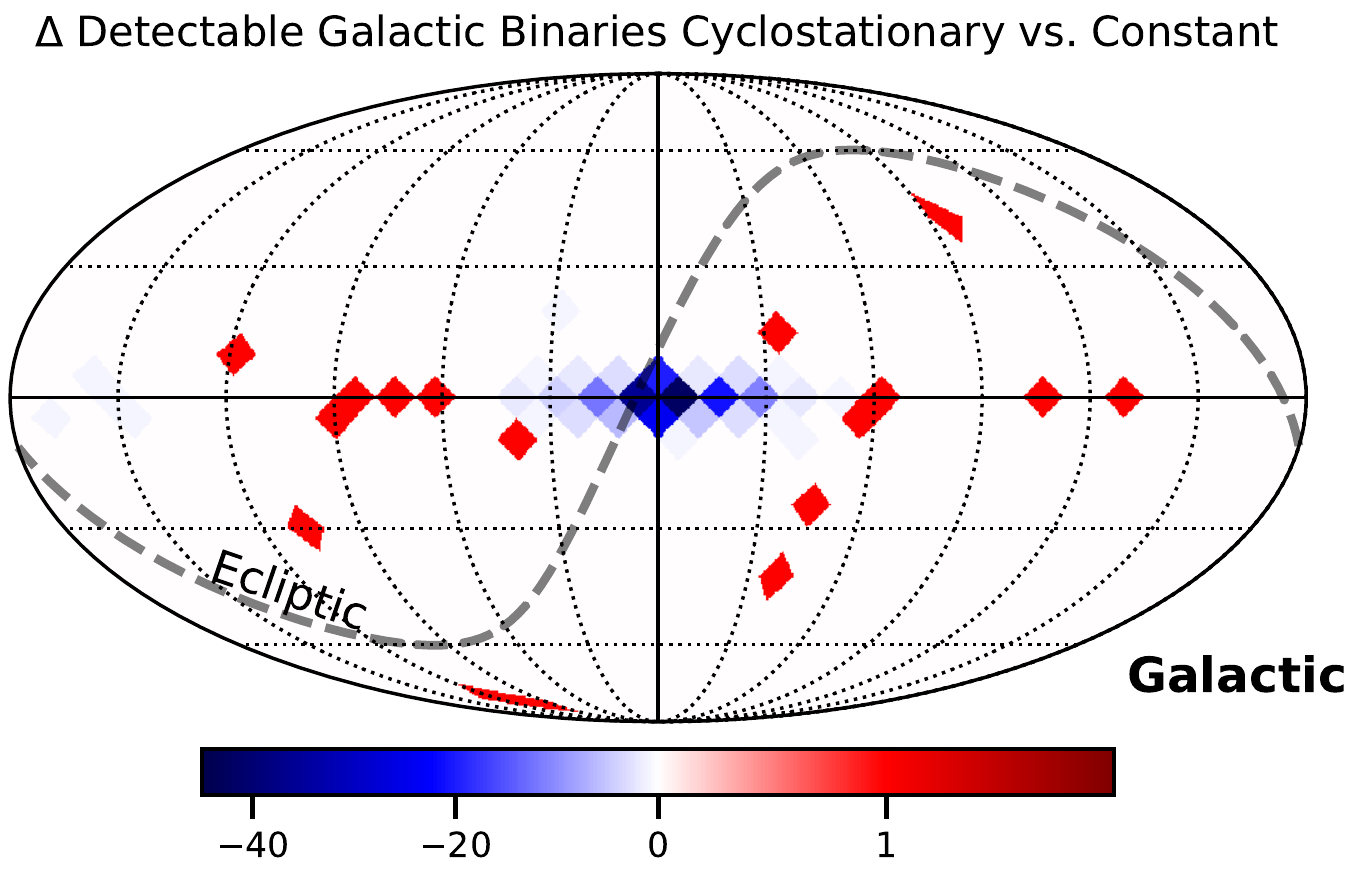}
\caption{\label{fig:gal_pos}Histogram comparison of the number of detectable galactic binaries between a constant model of the galactic stochastic background and cyclostationary model for our one-year simulated dataset. The models disagree on the threshold $\text{SNR}>7$ for 241 binaries, out of 7,273 total predicted detectable by the cyclostationary model, vs. 7,470 in the constant model. Because the binaries are physically highly clustered at the galactic center, the relative increase in detection efficiency far from the galactic center is significantly greater than the loss of efficiency near the galactic center. Therefore, the overall observing volume is larger in the cyclostationary model, despite the lower number of galactic binaries detected. The source of the observed structure is explored in Fig.~\ref{fig:gb_snr}.}
\end{figure}

\begin{figure}
\includegraphics[width=\columnwidth]{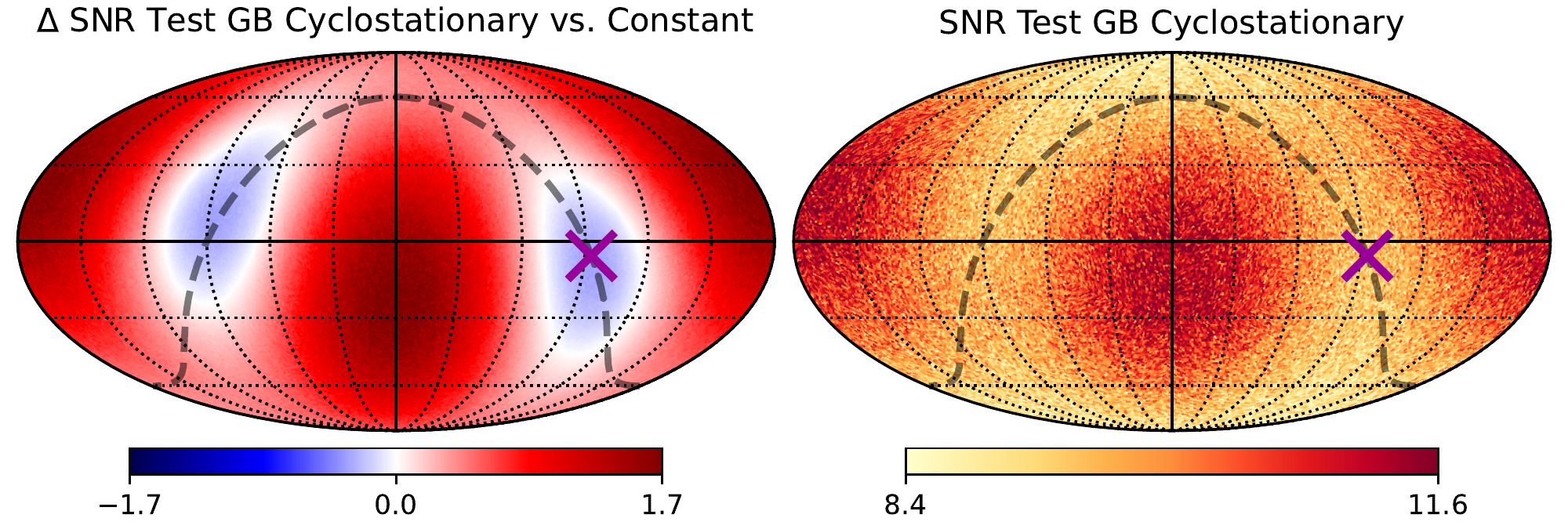}
\caption{Comparison of SNR obtained in our one year simulated dataset for a $2\;\text{mHz}$ galactic binary with all-sky-averaged $\text{SNR}\simeq10$. Each pixel is an average of 50 random realizations of the inclination, polarization, and binary phase at each HEALPix pixel. The total volume sensitive to the test binary is $\simeq20\%$ larger. However, the bulk of the galactic binaries predicted by the population synthesis model are found near the galactic center, resulting in the cyclostationary model detecting a lower number of galactic binaries, as shown in Fig.~\ref{fig:gal_pos}.}  \label{fig:gb_snr}
\end{figure}

In our simulated dataset, the constant model predicts slightly more galactic binaries are detectable than in our cyclostationary model. However, the individual detected binaries have a different distribution on the sky, as shown in Fig.~\ref{fig:gal_pos}. The difference in detectability is due to the interaction between LISA's antenna pattern and the concentration of binaries within the galaxy. As shown in Fig.~\ref{fig:gb_snr}, the cyclostationary model significantly improves sensitivity away from the galactic center and, to a lesser extent, degrades towards the galactic center and anticenter. Overall, the improved cyclostationary model generally predicts a larger total sensitive volume for LISA than the constant model by a factor that varies depending on the frequency and observation time, as shown in Table.~\ref{tab:bin_count}.

\begin{center}
\begin{table}
\begin{tabular}{||c c c c c c c c||} 
 \hline
 t (yrs) & $N_{\text{det,const}}$ & $N_{\text{det,cyclo}}$ & $N_{\text{disagree}}$ & $r_{\text{vol},1.0\;\text{mHz}}$ &$r_{\text{vol},1.5\;\text{mHz}}$ & $r_{\text{vol},2.0\;\text{mHz}}$ &$r_{\text{vol},2.5\;\text{mHz}}$\\ [0.5ex] 
 \hline\hline
 1 & 7470  & 7273  & 241 & 1.08 & 1.17 & 1.20 & 1.19 \\
 \hline
 2 & 11764 & 11512 & 298 & 1.08 & 1.16 & 1.21 & 1.05 \\
 \hline 
 3 & 15089 & 14831 & 308 & 1.11 & 1.04 & 1.06 & 1.10 \\
 \hline
 4 & 17992 & 17608 & 428 & 1.09 & 1.12 & 1.07 & 1.03 \\ 
 \hline
 5 & 20427 & 20018 & 477 & 1.10 & 1.09 & 1.08 & 0.99 \\ 
 \hline
 6 & 22674 & 22223 & 505 & 1.08 & 1.11 & 1.08 & 1.02 \\ 
 \hline 
 7 & 24854 & 24341 & 559 & 1.08 & 1.11 & 1.07 & 0.98 \\ 
 \hline
 8 & 26925 & 26417 & 576 & 1.07 & 1.08 & 1.03 & 1.02 \\
 \hline
\end{tabular}
\caption{\label{tab:bin_count}Detection efficiency for the binaries in the Sangria dataset as a function of total observation time, and the ratio of the sensitive observing volumes for an injected test binary at several different GW frequencies. At shorter observation times, the cyclostationary model detects less binaries near the galactic center but overall improves sensitivity. The cutoff frequency of the galactic background decreases over time, and the models begin to agree above the cutoff frequency, as shown in Fig.~\ref{fig:gb_evolve}.}
\end{table}
\end{center}

\subsection{Sensitivity to Supermassive Black Hole Binaries}\label{ssec:smbhb_sens}

Because LISA's antenna pattern rotates relative to the galaxy throughout the year, the degradation in LISA's sensitivity due to the galactic stochastic background also evolves with time. This effect is analogous to the obscuration of electromagnetic signals by the matter in the galactic plane. An excellent source class to illustrate the impact of the variation is supermassive black hole binaries, which evolve over timescales of days to weeks and therefore sample a different galactic background amplitude depending on their time of merger. 

\begin{figure}
\includegraphics[width=0.6\columnwidth]{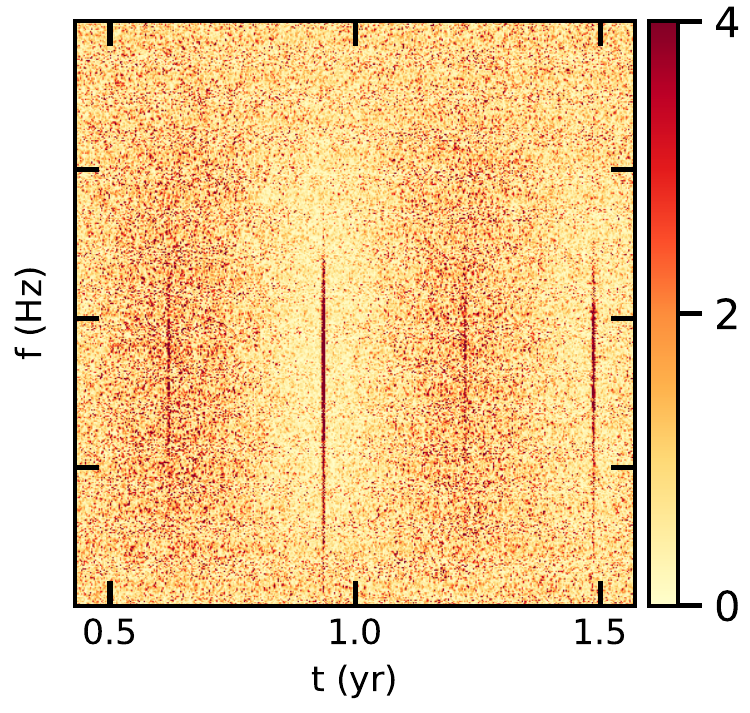}
\caption{Illustration of the injection procedure used to generate in Fig.~\ref{fig:smbhb_times} and Fig.~\ref{fig:smbhb_range}. For this figure, we superpose the simulated signals for an injected SMBHB at each of the four chirp times presented in Fig.~\ref{fig:smbhb_range} with a whitened galactic+instrument noise spectrum from the constant noise model. The source has a total mass chosen such that $f_{\text{RD}}=2\;\text{mHz}$, and a mean $\text{SNR}^2=10,000$ across all realizations of the chirp time and sky location.} \label{fig:chirp_illustration}
\end{figure}

\begin{figure}
\includegraphics[width=0.8\columnwidth]{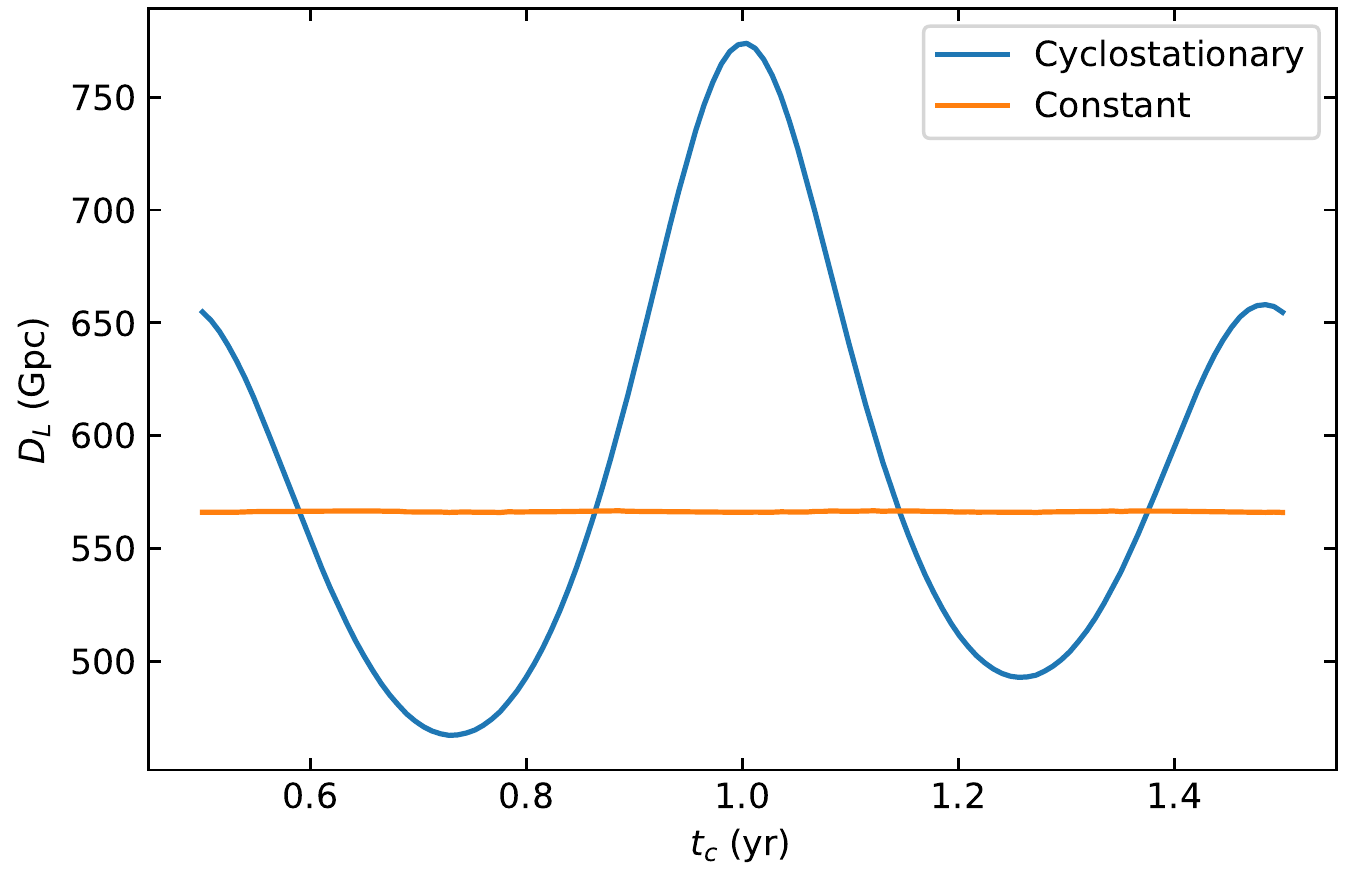}
\caption{Effective range to an injected supermassive black hole binary with $f_{RD}=2\;\text{mHz}$ as a function of chirp time using the galactic background from the two year simulated dataset. While the sensitivity varies with time, averaged over the course of a year the total observing volume is about 20\% larger. The minimum at $t_c\simeq0.73$ corresponds to the closest passage of LISA's maximum sensitivity to the galactic anticenter. }  \label{fig:smbhb_range}
\end{figure}

\begin{figure}
\includegraphics[width=\columnwidth]{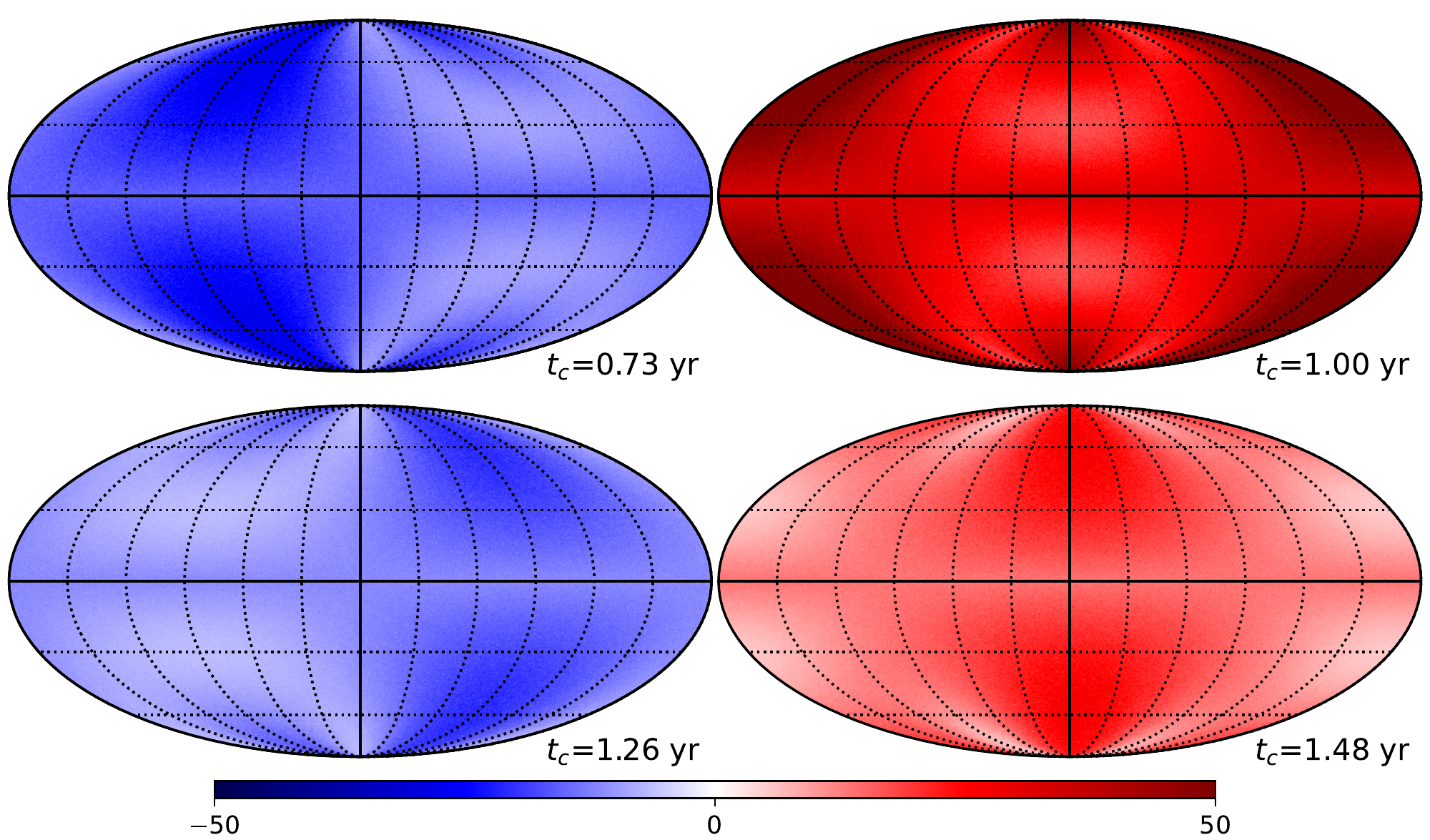}
\caption{Difference in predicted SNR between the cyclostationary and constant models for an injected $f_{RD}=2\;\text{mHz}$ supermassive black hole binary at four selected chirp times corresponding to the extrema in Fig.~\ref{fig:gb_evolve}. The minima in LISA's sensitivity to SMBHBs correspond to the times when LISA's peak sensitivity passing closest to the galactic center ($t_c\simeq1.26\;\text{yr}$) and closest to the galactic anticenter ($t_c\simeq0.73\;\text{yr}$). }\label{fig:smbhb_times}
\end{figure}

To illustrate the difference in sensitivity, we inject simulated SMBHB sources at a grid of chirp times $t_c$ and sky positions. The injection procedure as function of $t_c$ is shown in Fig.~\ref{fig:chirp_illustration}. In Fig.~\ref{fig:smbhb_times}, we show the impact of the cyclostationary model on LISA's sensitivity as a function of sky location at four selected chirp times. In Fig.~\ref{fig:smbhb_range}, we show the sky-averaged effective range of LISA as a function of sky location in the different models. 

As anticipated by the analogy with the electromagnetic case, LISA is less sensitive to mergers in the direction of the galactic center and anticenter than mergers that occur out of the galactic plane. The uneven distribution of noise on the sky over time will likely also impact parameter estimation for SMBHB sources even at fixed signal-to-noise ratio. For example, the shifting relative amplitude of the noise in the A and E channels may result in altered constraints on sky location, inclination, and polarization in some parts of the sky. There will also be parameter-space volume effects due to the larger sensitive volume away from the galactic plane, which will likely shift the mass range LISA is sensitive as a function of sky location. We will examine the effects of the time-varying galactic stochastic background on parameter estimation in more detail in a future publication.

\section{Conclusion}\label{sec:conclusion}

In this paper, we have extended existing iterative techniques for deriving the galactic stochastic background with LISA to handle the temporal variation in the galactic signal more correctly. Our technique significantly improves the quality of the fit of the noise model to the data. 

We have examined how the amplitude of the galactic stochastic background varies throughout the year as a function of sky location for various sources. We have also shown how the amplitude variation's level and shape change over the mission's lifetime. To facilitate future analyses of the effects of time-varying galactic stochastic background without replicating the entire iterative procedure, we have a simple empirical fit to the time variation in the galactic stochastic background in the ``Sangria'' dataset in Eq.~\ref{eq:r_periods}. We provide empirical coefficients for the annual cyclostationary time variation of the spectrum in Table~\ref{tab:evolve_params} and the spectral shape parameters include the secular dependence on observation time in Table~\ref{tab:fit_params}.

Overall, the improved modeling shows that results derived from the previous spectral models underestimate the total integrated observation volume of LISA by a factor of $\simeq10-20\%$ for sources that accumulate most of their SNR in the $1-3\;\text{mHz}$ frequency band. This improvement occurs because most of the sky is less severely contaminated than the galactic center. The level of improvement would depend on the specific galactic population model used, and it might be possible that some alternate galactic population models would not exhibit a significant improvement in the sensitive volume. We emphasize that our method is simply a more correct model of the noise spectrum, and it would still produce more correct results even for a choice of galaxy model where it predicted a lower total integrated observing volume. 

Future work will combine our code with an MCMC pipeline to investigate the impact of the improved galactic spectrum modeling on parameter estimation for various sources.The temporal variation in the galactic spectrum may have more significant implications for parameter estimation than expected based on its effect on the computation of SNR alone. Therefore, correctly modeling the temporal variation in the noise spectrum will be an essential component of pipelines to obtain a global fit to LISA data. 

Our technique can easily be extended to any cyclostationary noise source in LISA, or even arbitrary non-cyclostationary time-frequency noise models, provided the noise is adiabatic.  

Future analyses attempting to extract conclusions about LISA sources should use the improved noise modeling to more accurately model the impact of the galactic stochastic background on LISA data. 

\begin{acknowledgments}
This work was supported by NASA LISA foundation Science Grant 80NSSC19K0320.
\end{acknowledgments}

%






\appendix

\section{Appendix information}

\bibliography{nonstationary}{}
\bibliographystyle{aasjournal}

\end{document}